\newcommand{\LG}{\mathrm{LG}}
\newcommand{\HG}{\mathrm{HG}}

\documentclass[pra,reprint,showpacs,amsmath,amssymb,superscriptaddress,eqsecnum,longbibliography]{revtex4-1}
\usepackage{graphicx,bm,color,mathptmx,hyperref}

\begin{document}

\title{Experimental violation of a Bell-like inequality with optical vortex beams}

\author{B. Stoklasa}
\affiliation{Department of Optics, 
Palack\'y  University, 17. listopadu 12, 
771 46 Olomouc, Czech Republic}

\author{L. Motka}
\affiliation{Department of Optics, 
Palack\'y  University, 17. listopadu 12, 
771 46 Olomouc, Czech Republic}

\author{J. Rehacek}
\affiliation{Department of Optics, 
Palack\'y  University, 17. listopadu 12, 
771 46 Olomouc, Czech Republic}

\author{Z. Hradil}
\affiliation{Department of Optics, 
Palack\'y  University, 17. listopadu 12, 
771 46 Olomouc, Czech Republic}

\author{L. L. S\'{a}nchez-Soto}
\affiliation{Departamento de \'Optica,
  Facultad de F\'{\i}sica, Universidad Complutense, 
28040~Madrid,   Spain}
\affiliation{Max-Planck-Institut f\"ur die Physik des Lichts,
  G\"{u}nther-Scharowsky-Stra{\ss}e 1, Bau 24, 91058 Erlangen,
  Germany}

\author{G. S. Agarwal}
\affiliation{Department of Physics, Oklahoma
  State University, Stillwater, Oklahoma 74078, USA}

\begin{abstract}
  Optical beams with topological singularities have a Schmidt
  decomposition.  Hence, they display features typically associated
  with bipartite quantum systems; in particular, these classical
  beams can exhibit entanglement.  This classical entanglement can be
  quantified by a Bell inequality formulated in terms of Wigner
  functions.  We experimentally demonstrate the violation of this
  inequality for Laguerre-Gauss (LG) beams and confirm that
  the violation increases with increasing orbital angular momentum.
  Our measurements yield negativity of the Wigner function
  at the origin for $\LG_{10}$ beams, whereas for $\LG_{20}$ we always
  get a positive value.
\end{abstract}


\maketitle

\section{Introduction}

Entanglement is usually presented as one of the weirdest features of
quantum theory that depart strongly from our common
sense~\cite{Schrodinger:1935ys}.  Since the seminal work of Einstein,
Podolsky, and Rosen (EPR)~\cite{Einstein:1935yt}, countless
discussions on this subject have
popped up~\cite{Horodecki:2009vu}.

A major step in the right direction is due to Bell~\cite{Bell:1964gc}, who
formulated the EPR dilemma in terms of an inequality which naturally
led to a falsifiable prediction. Actually,  it is common to use
an alternative formulation, derived by  Clauser, Horne, Shimony and
Holt (CHSH)~\cite{Clauser:1969it}, which is better suited for realistic
experiments.

The main stream of research~\cite{Brunner:2014ys,Werner:2001fj} settled
the main concepts of this topic in the realm of quantum
physics. However, in recent years a general consensus has been reached
on the fact that entanglement is not necessarily a signature of the
quantumness of a system. Actually, as aptly remarked in
RefE.~\cite{Toppel:2014jt}, one should distinguish between two types of
entanglement: between spatially separated systems (inter-system
entanglement) and between different degrees of freedom of a single
system (intra-system entanglement). Inter-system entanglement occurs
only in truly quantum systems and may yield to nonlocal statistical
correlations.  Conversely, intra-system entanglement may also appear
in classical systems and cannot generate nonlocal
correlations~\cite{Brunner:2005gv}; for this reason, it is often
dubbed as ``classical entanglement''. Since its introduction by
Spreeuw~\cite{Spreeuw:1998ho}, this notion has been employed in a
variety of contexts~\cite{Ghose:2014oe}. 
 
Classical entanglement has allowed to test Bell inequalities with
classical wave fields. The physical significance of this violation is
not linked to quantum nonlocality, but rather points to the
impossibility of constructing such a beam using other beams with
uncoupled degrees of freedom.  However, all the experiments conducted
thus far to observe this violation have involved only discrete
variables, such as spin and beam path of single
neutrons~\cite{Hasegawa:2003hf}, polarization and transverse modes of
a laser beam~\cite{Souza:2007fb,Simon:2010jk,Qian:2011jc,
  Gabriel:2011bn,Eberly:2014}, different transverse modes propagating
in multimode waveguides~\cite{Fu:2004fk}, polarization of two
classical fields with different frequencies~\cite{Lee:2002uq}, orbital
angular momentum~\cite{Goyal:2013la,Chowdhury:2013}, and polarization
and spatial parity~\cite{Kagalwala:2013br}.

In this paper, we continue the analysis of this classical entanglement
by focusing on the simple but engaging example of vortex beams. To
this end, in Sec.~\ref{sec:Schmidt} we revisit a decomposition of
Laguerre-Gauss (LG) beams in the Hermite-Gauss (HG) basis that can be
rightly interpreted as a Schmidt decomposition. This immediately
suggests that many ideas ensuing from the quantum world may be
applicable to these beams as well. In particular, in
Sec.~\ref{sec:CHSH} we address the inseparability of the LG modes
using a CHSH violation that we quantify in terms of the associated
Wigner function. As this distribution can be understood as a measure
of the displaced parity, in Sec.~\ref{sec:exp} we discuss an
experimental realization which nicely agrees with the theoretical
predictions. Finally, our conclusions are summarized in
Sec.~\ref{sec:concl}.

\section{Optical vortices and Schmidt decomposition}
\label{sec:Schmidt}

It is well known that the beam propagation along the $z$ direction of
a monocromatic scalar field of frequency $\omega$; i.e.,
$E (\mathbf{r}, t) = \mathcal{E} (\mathbf{r} ) \exp [- i ( \omega t -
k z )] $, is governed by the paraxial wave equation
\begin{equation}
  \frac{\partial \mathcal{E}}{\partial z}  = -\frac{\lambdabar}{2}
  \left( \frac{\partial^2}{\partial x^2} +
    \frac{\partial^2}{\partial y^2}\right) \mathcal{E} \, ,
  \label{freespace}
\end{equation}
with $\lambdabar = \lambda/2\pi$ and $\lambda $ is the
wavelength. Equation~(\ref{freespace}) is formally identical to the
Schr\"{o}dinger equation for a free particle in two dimensions, with
the obvious identifications $t \mapsto z$, $\psi \mapsto \mathcal{E}$,
and $\hbar \mapsto \lambdabar$.

Any optical beam can be thus expressed as a superposition of
fundamental solutions of Eq.~(\ref{freespace}).  In Cartesian
coordinates, a natural orthonormal set is given by the Hermite-Gauss
(HG) modes:
\begin{gather}
  \HG_{mn} ( x, y ) = \sqrt{\frac{2}{\pi n! m! 2^{n+m}}} \left (
    \frac{1}{w} \right ) H_{m} \left ( \frac{\sqrt{2} x}{w} \right )
  H_{n }  \left ( \frac{\sqrt{2} y}{w} \right ) \nonumber \\
  \times \exp \left ( - \frac{x^{2} + y^{2}}{w^{2}} \right ) ,
  \label{waveHG}
\end{gather}
where $w$ is the beam waist, and $H_{m}$ are the Hermite
polynomials. Note that we are restricting ourselves to the plane
$z=0$, since we are not interested here in the evolution.

For cylindrical symmetry, it is convenient to use the set of
Laguerre-Gauss (LG) modes, which contain optical vortices with
topological singularities; they read
\begin{gather}
  \LG_{mn} ( r,\varphi ) = \sqrt{\frac{2}{\pi m! n!}}  \min (m,n)!
  (-1)^{\min (m,n)}
  \left ( \frac{1}{w} \right )  \nonumber \\
  \times \left ( \frac{\sqrt{2} r}{w} \right )^{|m-n|} \!\!  L_{\min
    (m,n)}^{|m - n|} \left ( \frac{2 r^{2}}{w^{2}} \right ) \exp \left
    (- \frac{r^{2}}{w^{2}} \right ) \exp [ i (m-n) \varphi ] \, ,
  \label{waveLG}
\end{gather}
where $L_{p}^{|\ell |} (x)$ are the generalized Laguerre
polynomials. A word of caution seems to be in order: usually, these
modes are presented in terms of two different indices: the azimuthal
mode index $\ell = m - n$, which is a topological charge giving the
number of $2 \pi$-phase cycles around the mode circumference, and
$p = \min (m, n)$ is the radial mode index, which is related to the
number of radial nodes~\cite{Karimi:2014yu}. However, the form
(\ref{waveLG}) will be advantageous in what follows.

The crucial observation is that the LG  modes can be represented as
superpositions of HG modes, and viceversa. This can be compactly
written down as~\cite{Beijersbergen:1993eu}
\begin{equation}
  \LG_{mn} (\rho,\varphi) = \sum_{k=0}^{m+n}  B_{mn}^{k}  \,
\HG_{m+n-k,k} (x,y)
  \label{legherm}
\end{equation}
where the coefficients are
\begin{equation}
B_{mn}^{k} =  \sqrt{\frac{k! (m +n-k)!}{m! n! 2^{n+m}}} \frac{(-i)^{k}}{k!}
\frac{d^k}{dt^k}  \left . [ (1-t)^m(1+t)^n]  \right |_{t=0} .
  \label{def11}
\end{equation}
This looks exactly the same as a Schmidt decomposition for a bipartite
quantum system.  It is nothing but a particular way of expressing a
vector in the tensor product of two inner product
spaces~\cite{Peres:1993fk}. Alternatively, it can be seen as another
form of the singular-value decomposition~\cite{Stewart:1993uq}, which
identifies the maximal correlation directly. In quantum
information, the Schmidt coefficients $B_{mn}^{k}$ convey complete
information of the entanglement~\cite{Agarwal:02}. Here, we intend to 
 assess entanglement in LG beams via the violation of suitably
 formulated Bell inequalities. 

\section{CHSH violation for Laguerre-Gauss modes}
\label{sec:CHSH}

The traditional form of the CHSH inequality applies to dichotomic
discrete variables.  For continuous variables, the sensible
formulation is in terms of the Wigner function,
which for a classical beam reads
\begin{equation}
  W(\mathbf{x},\mathbf{p}) = \frac{1}{\lambdabar^2 \pi^2}
  \int d^2\mathbf{x}^{\prime} \,  
   e^{2i \mathbf{p} \cdot \mathbf{x}^{\prime}/\lambdabar}
  \langle E^{\ast}(\mathbf{x} - \mathbf{x}^{\prime}) E(\mathbf{x} +
  \mathbf{x}^{\prime}) \rangle \, ,
  \label{wigel}
\end{equation}
the angular brackets denoting statistical average. Although originally
introduced to represent quantum mechanical phenomena in phase
space~\cite{Wigner:1932kx}, the Wigner distribution was established in
optics~\cite{Walther:1968qy} to relate partial coherence with
radiometry. Since then, a great number of applications of this
function have been reported~\cite{Bastiaans:2009sp,
  Galleani:2002hb,Dragoman:1997rw,Mecklenbrauker:1997tx,Alonso:2011pi}.
Note that $W$ has the dimensions of an intensity and it yields a
description displaying both the position and the momentum (which in
the paraxial approximation has the significance of a scaled angular
coordinate) of the intensity of the wave field: in fact, one easily
proves that
\begin{eqnarray}
 &\displaystyle
  \int   W(\mathbf{x},\mathbf{p})     \, d\mathbf{p} = 
 I ( \mathbf{x} ) \equiv \langle E^{\ast} (\mathbf{x} )
  E(\mathbf{x} \rangle \, , & \nonumber \\
& & \\
& \displaystyle
 \frac{1}{\lambdabar^2 \pi^2}  \int   W(\mathbf{x},\mathbf{p})     \,
  d\mathbf{x}  = 
  I (  \mathbf{p} ) \equiv 
  \langle E^{\ast} (\mathbf{p} ) E (\mathbf{p}) \rangle \, , & 
\nonumber
\end{eqnarray}
with 
\begin{equation}
E ( \mathbf{p} )=  \frac{1}{\lambdabar^2 \pi^2}  
\int  E (  \mathbf{x} ) \, \exp( i \mathbf{p} \cdot
\mathbf{x}/\lambdabar)  \, d\mathbf{x} \, .
\end{equation}
 Thus, the marginals of the Wigner function  are the intensity
 distributions in $\mathbf{x}$ or $\mathbf{p}$ space, respectively.

The CHSH inequality can now be stated in terms of the Wigner function 
 as~\cite{Banaszek:1999mw}
\begin{equation}
  B =  \frac{\pi^2}{4}  | W (\alpha, \beta) +
  W(\alpha,\beta^{\prime}) + W ( \alpha^{\prime},\beta)
  - W(\alpha^{\prime},\beta^{\prime})| < 2,
  \label{belin12}
\end{equation}
where $\alpha = (x, p_{x} )/\sqrt{2}$ and $\beta = (y,
p_{y})/\sqrt{2}$. This also follows from
the work of Gisin~\cite{Gisin1991201}, who formulated a Bell inequality for the
set of observables with the property $\hat{O}^{2} = \openone$: as we
shall see, the Wigner function appears as the average
value of the parity, whose square is unity. Reference~\cite{Chowdhury:2013}
presents a detailed study of the violations of (\ref{belin12}).

For the state $\LG_{mn}$, the normalized
Wigner function can be written as~\cite{Simon:2000xp}
\begin{gather}
  W_{mn}^{\LG} ( X, P_{X}; Y, P_{Y} )  =  \frac{(-1)^{m+n}} {\pi^{2}}
  \exp(-4 Q_0)  \nonumber \\
 \times  L_{m} [4  (Q_0+Q_2)]  L_{n} [4(Q_0-Q_2)] \, ,
  \label{WF_LG_n1_n2}
\end{gather}
where 
\begin{equation}
Q_{0} =  \frac{1}{4} ( X^{2} + Y^{2} + P_{X}^{2} + P_{Y}^{2} ) \, ,
\qquad 
Q_{2}  =  \frac{1}{2} ( X P_{Y} - Y P_{X} ) \, ,
\end{equation}
and  we have rescaled the variables as $ x  \mapsto (w/\sqrt{2}) X$
and  $p_x  \mapsto (\sqrt{2} \lambdabar/w)~P_X$ (and analogously
for the $y$ axis).  Let us first look at the simple case
of the mode $\LG_{10}$,  which reduces to
\begin{gather}
  W^{\LG}_{10} (X, P_X; Y, P_Y) =  \frac{1}{\pi^2}   \exp(
  -P_X^2-P_Y^2-X^2-Y^2 )  \nonumber \\
\times [ (P_X - Y)^2 + (P_Y+X)^2 -1 ] \, .
\end{gather}
 The two measurement settings on one side are chosen to be
$\alpha = (  X = 0, P_{X}=0 ) $ and  $ \alpha^{\prime} = (
X^{\prime} = X,  P_{X}^{\prime}=0 )$, and the corresponding
settings on the other side are $\beta = ( Y=0, P_{Y} = 0 ) $
and $\beta^{\prime} = (Y^{\prime}=0, P_{Y}^{\prime} =
P_{Y})$~\cite{Zhang:2007xb}, for which the Bell sum is
\begin{gather}
  B =  e^{-P_Y^2}  (P_Y^2 - 1 ) + e^{-X^2}  ( X^2 - 1 )   \nonumber \\
   -  e^{-( P_Y^2 + X^2)} [ ( P_Y + X)^2 - 1]  -1 \, .
       \label{BI_1_3}
\end{gather}
Upon maximization with respect to $X$ and $P_{Y}$, we obtain the
maximum Bell violation, $|B_{\mathrm{max}}| \simeq 2.17$, which
happens for the choices
$X \simeq 0.45,~P_{Y} \simeq 0.45$~\cite{Chowdhury:2013}.  For
comparison, note that the maximum Bell violation in quantum mechanics
through the Wigner function for the two-mode squeezed vacuum state
using similar settings is given by
$|B_{\mathrm{max}}^{\mathrm{QM}} | \simeq 2.19$~\cite{Banaszek:1999mw}.

The Bell violation may be further optimized by a more general choice
of settings than those used here.  For example, maximizing it with
respect to the parameters $\alpha = (X,P_{X})$,
$\alpha^{\prime}= (X^{\prime}, P_{X}^{\prime})$, $\beta= (Y, P_{Y})$,
$\beta^{\prime} = (Y^{\prime}, P_{Y}^{\prime})$, one obtains the
absolute maximum Bell violation, $|B_{\max}| = 2.24$ and occurs for
the choices $X \simeq -0.07,~P_{X} \simeq 0.05,~X^{\prime} \simeq
0.4,~P_{X}^{\prime} \simeq-0.26,~Y \simeq-0.05,~P_{Y} \simeq
-0.07,~Y^{\prime} \simeq 0.26,~P_{Y}^{\prime} \simeq 0.4$.
The violation also increases with higher orbital angular
momentum. This increase with $n$ is analogous to the enhancement of
nonlocality in quantum mechanics for many-particle
Greenberger-Horne-Zeilinger states~\cite{Mermin:1990nf}.

\section{Experimental results}
\label{sec:exp}

\begin{figure}[b]
  \centerline{\includegraphics[width=0.95\columnwidth]{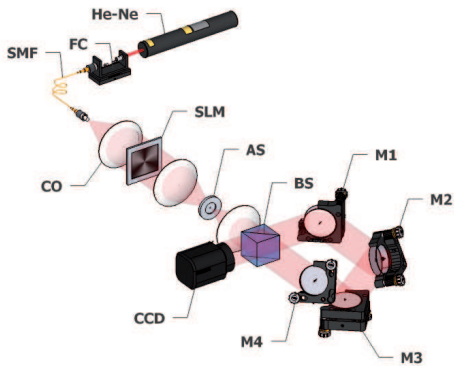}}
  \caption{(Color online) Scheme of the Bell measurement. The
    abbreviations are as follows: He-Ne: laser source, FC: fiber
    coupler, SMF: single mode fiber, CO: collimation optics, SLM:
    spatial light modulator, AS: aperture stop, BS: beam splitter,
    M1-M4: mirrors, CCD: camera}
  \label{figSetup}
\end{figure}

We have carried a direct measurement of the Bell sums for optical
beams with different amount of nonlocal correlations.  To understand
the measurement, we recall that the Wigner function in quantum optics
is often regarded as the average of the displaced parity
operator~\cite{Royer:1977qf}. At the classical level, we can consider
the field amplitudes $\mathcal{E} (X, Y)$ as vectors in the Hilbert
space of complex-valued functions that are square integrable over a
transverse plane. In this space we define linear Hermitian operators
\begin{equation}
  \hat{X}: \mathcal{E}  (X, Y )
  \mapsto X \mathcal{E} (X, Y )   \, ,
  \qquad
  \hat{P}_{x}: \mathcal{E}  (X, Y )
  \mapsto -  i  \frac{\partial}{\partial X}
   \mathcal{E}  (X, Y )   \, ,
\end{equation}
and analogous ones for the $Y$ variable. Formally, these operators
satisfy the canonical commutation relations
$ [\hat{X}, \hat{P}_{X} ] = [\hat{Y}, \hat{P}_{Y} ] = i $. Therefore,
the unitary parity operator is 
\begin{equation}
\hat{\Pi}_{X} \,
\hat{X} \, \hat{\Pi}_{X} = - \hat{X}   \, , 
\qquad
\hat{\Pi}_{X} \, 
\hat{P}_{X} \,\hat{\Pi}_{X} = - \hat{P}_{X} \, ,
\end{equation}
and changes $\mathcal{E} (X, Y)$ into $\mathcal{E} (-X, Y)$.  The
displacement operators are  
\begin{equation}
\hat{D} (X, P_{X}) =
\exp [ i (P_{X} \hat{X} - X \hat{P}_{X} )] \, . 
\end{equation}
Indeed,  with these notations we have
\begin{align}
  W (X, P_{x}; Y, P_{y} )  & =  \nonumber \\
&   \frac{4}{\pi^2}
 \langle
  \hat{D} (X, P_{x}) \hat{\Pi}_{X} \hat{D}^{\dagger}(X, P_{x})
\,  \hat{D} (Y, P_{y}) \hat{\Pi}_{Y} \hat{D}^{\dagger}(Y, P_{y})
  \rangle \, .
\end{align}

\begin{figure}
  \centerline{\includegraphics[width=0.97\columnwidth]{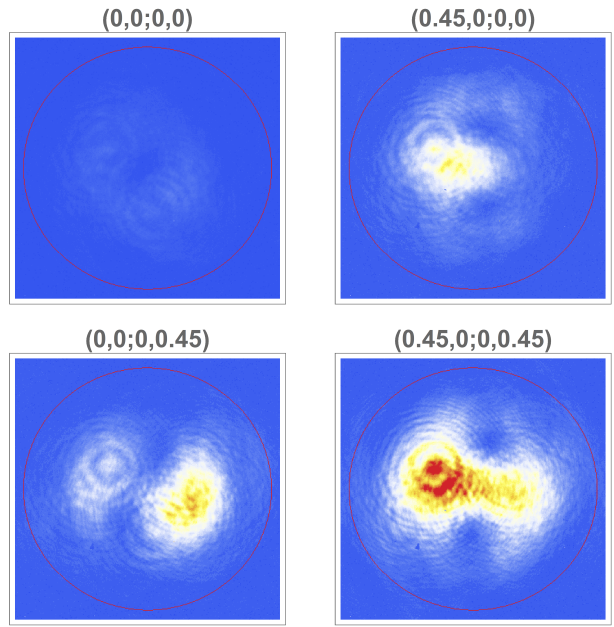}}
  \caption{(Color online) Snapshots of the CCD camera for the state
    $\LG_{10}$ at the four settings $(X, P_{X}; Y, P_{Y})$ indicated. The scans are
    normalized to the peak intensity among the measurements and the
    area of interest for the intensity integration is marked by a red
    circle.}
  \label{figCCD}
\end{figure}

Parity measurement can be, in turn, realized by a common-path
interferometer with a Dove prism inserted into the optical
path~\cite{Mukamel:2003qq}. In our setup, sketched in
Fig.~\ref{figSetup}, the prism was substituted with an equivalent
four-mirror Sagnac arrangement~\cite{Smith:2005vo}.  The two copies of
the input signal obtained after the input beam splitter are
transformed by the mirrors so as to make one copy spatially inverted
with respect to the other, prior to combining the beams together. The
resulting interference pattern is detected by a CCD camera:
Figure~\ref{figCCD} shows snapshots of the camera for the state
$\LG_{10}$ at the four settings indicated.  The total intensity
witnessing parity of the measured beam is computed by spatial
integration and this is proportional to the desired Wigner
distribution sample after normalization to the overall intensity.

The target signal beams were prepared with digital holograms created
by a spatial light modulator (SLM), which modulated a collimated
output of a single mode fiber coupled to a He-Ne laser. We also
included a 4$f$-system, with an aperture stop, to filter the unwanted
diffraction orders produced by the SLM. To allow for a better
flexibility, all the necessary shifts in the $X$, $Y$, $P_{X}$, and
$P_{Y}$ variables were incorporated into the SLM, so that each
Bell measurement was associated with a separate hologram.

\begin{figure}
  \centerline{\includegraphics[width=\columnwidth]{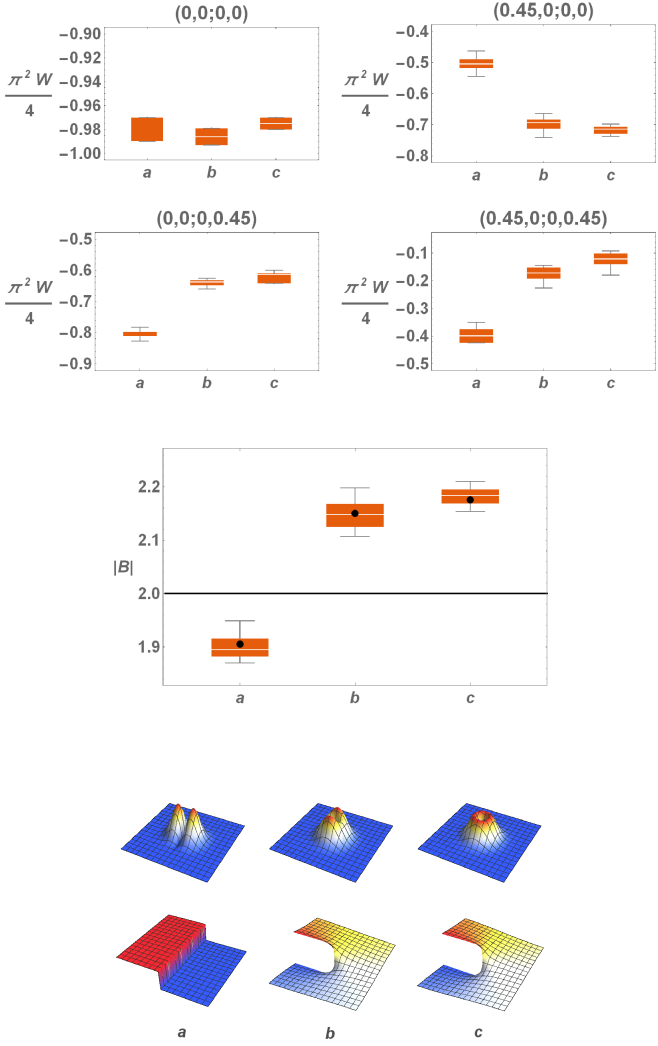}}
  \caption{(Color online) Experimental results for three different
    optical beams: a) $\HG_{10}$, b)
    $0.4 \, \HG_{10}+ i 0.6 \, \HG_{01}$, and c) $\LG_{10}$.  At the
    top, we plot $\frac{\pi^{2}}{4} W(X, P_{X}; Y, P_{Y})$ at the
    values $(X,P_{X}; Y, P_{Y})$ indicated for each one.  The next
    plot shows the measured Bell sums, all reported with $75\%$ and
    $25\%$ quartile (orange boxes) and the minimal and maximal
    measured values (error bars). The theoretical values
    $(-1.91, -2.15, -2.17)$ are the dots and the black bar is at
    $|B|=2$, which delimites the classically entangled states.  The
    theoretical amplitude (top) and phase (bottom) distributions of
    the measured beams are plotted bellow the chart.}
  \label{figResults}
\end{figure}

The measured beams were coherent superpositions of Hermite-Gaussian
beams in the form $a \, \HG_{10}+ i b \, \HG_{01}$ with
$\{a=1,\, b=0\}$, $\{a=0.4,\, b=0.6\}$ and $\{a=0.5,\, b=0.5\}$,
respectively. The first and the third are thus a pure Hermite-Gaussian
beam and a pure Laguerre-Gaussian vortex beam, respectively.  For all
the beams we used the settings
$X \simeq 0.0,~P_{X} \simeq 0.0,~X^{\prime} \simeq
-0.45,~P_{X}^{\prime} \simeq 0.0,~Y \simeq 0.0,~P_{Y} \simeq
0.0,~Y^{\prime} \simeq 0.0,~P_{Y}^{\prime} \simeq -0.45$
for the evaluation of the Bell sums. The theoretical values of the
Bell sums for these are $ (-1.91, -2.15, -2.17)$, respectively.

Each measurement was repeated many times with slightly different
readings, due to laser intensity instabilities and CCD noise. These
effects manifest as measurement errors, which can be estimated from
the sample statistics. As the parity measurement requires to normalize
the total measured intensity of the interference pattern with respect
to the input beam intensity, a separate reading of the input beam
intensity was performed. For each optical beam, the mean value of the
Bell sum is reported. The results are summarized in
Fig.~\ref{figResults}. The Bell correlations grow with the coupling
between the basis $\HG_{10}$ and $\HG_{01}$ modes, with statistically
significant violation of CHSH inequality by the second and third
beams, as theoretically predicted.

 We also show the measured values of the Wigner function. For
 both, $\HG_{10}$ and $\LG_{10}$ modes,  the values of
 $\pi^2 W (0, 0; 0, 0)$ are quite close to $-1$. For classical
 beams, ours is one of the few measurements on the negativity of the
 Wigner function, though it has to be anticipated from the
 corresponding results in quantum optics~\cite{Schleich:2000}. We note
 that very early, March and Wolf~\cite{Wolf:1974} had constructed
 an example of a classical source which exhibited negative Wigner
 function.

\begin{figure}
  \centerline{\includegraphics[width=\columnwidth]{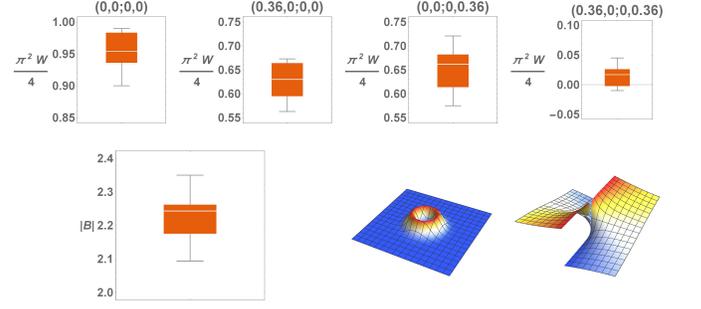}}
  \caption{(Color online) Experimental results for the beam
    $\LG_{20}$, as presented in Fig.~3. The values of $(X,0;0, P_{Y})$
    are also indicated. The Bell sum variation is significantly larger
    than in the case of $LG_{10}$. The plots on the right bottom panel
    are the amplitude and phase of $\LG_{20}$.}
  \label{figResults20}
\end{figure}

 Finally, we have checked the violation of CHSH inequality for the
 beam $\LG_{20}$. A beam with higher topological charge is more
 sensitive to setup imperfections, hence the Bell sum variation is
 significantly larger than in the case of $\LG_{10}$. On the other
 hand, as shown in Fig.~\ref{figResults20}, the increasing of the Bell
 sum for higher orbital angular momentum is clearly demonstrated:
 the theoretical value for $\LG_{20}$ is $-2.24$, which agrees pretty
 well with  the experimental results.
 \footnote{A study of the Bell violations for LG beams is also presented by
   S. Prabhakar, S. G. Reddy, A. A. Chithrabhanu, P. G. K. Samantha, and
   R. P. Singh, arxiv:1406.6239, although the authors does not employ
   parity measurements, but Fourier transform.}. 
Note that the Wigner function at the origin for  the $\LG_{20}$ beam
is positive, as expected. 

\section{Concluding remarks}
\label{sec:concl}

In short, we have presented an experimental study of nonlocal
correlations in classical  beams with topological
singularities~\cite{Chowdhury:2013}. These correlations between modes are
manifested through the violation of a CHSH inequality, which we have
detected via direct parity measurements. Such a violation is shown to increase with the value of
orbital angular momentum of the beam. As a byproduct of our
measurements, we obtain negativity of the Wigner function at certain
points in phase space for the $\HG_{10}$ and $\LG_{10}$ beams. Note
that this has implications for similar studies with electron beams,
for which vortices have been reported~\cite{Schattschneider:2010,Unguris:2011}.

Though entanglement here does not bear any paradoxical meaning, such
as ``spooky action on the distance'',  it  still represents a
potential resource for classical signal processing.
It might be expected  that  future applications of quantum information
processing can be tailored in terms of  classical light: the research
presented in this work  explores one of those options.

Furthermore, our results are relevant not only for a correct understanding
of ``classical entanglement'', but also for bringing out different
statistical features of the optical beams, since it provides an
alternative paradigm to the well developed optical coherence theory.

\begin{acknowledgments}
We acknowledge illuminating discussions with Gerd Leuchs, Elisabeth
Giacobino, and Andrea Aiello. This work was supported by the Grant
Agency of the Czech Republic (Grant 15-031945), the European Social
Fund and the State Budget of the Czech Republic POSTUP II (Grant
CZ.1.07/2.3.00/30.0041), the IGA of the Palack\'y University (Grant
PrF-2015-002), the Spanish MINECO (Grant FIS2011-26786), and UCM-Banco
Santander Program (Grant GR3/14).
\end{acknowledgments}


%

\end{document}